\documentclass[11pt]{article}
\hoffset = - 0.5 truecm
\usepackage{amssymb}
\usepackage{amsmath}
\usepackage[dvipsnames]{xcolor}
\usepackage{ulem,lipsum}
\usepackage{multirow}
\def\noi{\noindent}
\def\lsim{\raise0.3ex\hbox{$<$\kern-0.75em\raise-1.1ex\hbox{$\sim$}}}
\def\gsim{\raise0.3ex\hbox{$>$\kern-0.75em\raise-1.1ex\hbox{$\sim$}}}
\hoffset = - 0.5 truecm 
\def\beq{\begin{equation}}   \def\eeq{\end{equation}}
\def\noi{\noindent} \def\beeq{\begin{eqnarray}}
\def\eeeq{\end{eqnarray}}
\def\bea{\begin{eqnarray}}  \def\eea{\end{eqnarray}} \def\nn{\nonumber}
\newcommand\mysection{\setcounter{equation}{0}\section}

\renewcommand{\theequation}{\thesection.\arabic{equation}}

\newcounter{hran} \renewcommand{\thehran}{\thesection.\arabic{hran}}

\def\bmini{\setcounter{hran}{\value{equation}}

\refstepcounter{hran}\setcounter{equation}{0}

\renewcommand{\theequation}{\thehran\alph{equation}}\begin{eqnarray}}

\def\bminiG#1{\setcounter{hran}{\value{equation}}

\refstepcounter{hran}\setcounter{equation}{-1}

\renewcommand{\theequation}{\thehran\alph{equation}}

\refstepcounter{equation}\label{#1}\begin{eqnarray}}

\def\emini{\end{eqnarray}\relax\setcounter{equation}{\value{hran}}\renewcommand{\theequation}{\thesection.\arabic{equation}}}

\begin{document}

\pagestyle{plain}

\baselineskip 18pt

\begin{titlepage}

\vspace{1.cm}

\long\def\symbolfootnote[#1]#2{\begingroup%
\def\thefootnote{\fnsymbol{footnote}}\footnote[#1]{#2}\endgroup}

\begin{center}

{\large \bf Isolated photon cross section and resummation of the logarithms of the cone radius.}\\[2cm]

{\large  M.~Fontannaz$^{a}$, J.~Ph.~Guillet$^{b}$ } \\[.5cm]

\normalsize
{$^{a}$ Université Paris-Saclay, CNRS/IN2P3, IJCLAB, Pôle Théorie, F-91405 Orsay cedex, France}\\
{$^{b}$ LAPTH, Univ. Savoie Mont Blanc, CNRS, F-74000 Annecy, France}\\

\today
\end{center}

\vspace{2cm}

\begin{abstract}
\noindent
The cone isolation criterion used in large-pt photon experiments generates logarithms of the cone radius in the theoretical cross sections. When a small radius is used, unitarity is violated as the inclusive cross section is smaller than the isolated one. We show that unitarity is restored when these logarithms are resummed. We also study a criterion which offers a more precise description of the experimental procedure : no isolation is imposed in a very small cone inside the standard one. In this case as well unitarity is violated and the resummation of logarithms is required.
\end{abstract}

\vspace{1cm}


\vspace{2cm}

\end{titlepage}

\newpage

\mysection{Introduction}
\hspace*{\parindent} 
The detection of large-$p_T$ prompt\footnote{By prompt we mean not coming from a hadron decay
($\pi^0 \to \gamma\gamma, \eta \to \gamma\gamma$, etc .)} photons in hadronic collisions is made
difficult by the presence of numerous large-$p_T$ $\pi^0$ which decay in two photons. These $\pi^0$
belong to jets and are accompanied by  hadronic energy. Therefore the contributions to the
cross-section of fake ``$\pi^0$-photon'' can be strongly suppressed by imposing a cut $E_{T \, \text{cut}}$
on the transverse energy which accompanies the photons. For instance the cone-criterion imposes a
cut on the hadronic energy detected in a cone of radius $R$, in rapidity and azimuthal angle, drawn
around the observed photon. 
The cone-criterion is used in theoretical calculations but it does not correspond to the experimental  procedure. 
Usually, experiments do not impose cuts in a very small cone around the photon but rather in a ring centered in the same direction. 
For instance, in Atlas experiments \cite{1r} the transverse energy is the energy measured between a small cone of radius $r$ and a large cone of radius $R$. 
It is obtained by subtracting from the large cone energy, including the photon energy, the one measured inside the small cone. 
A similar criterium is used by the CMS collaboration \cite{2r}. We name this criterion hollow-cone isolation.

Of course these criteria have an impact on the sample of prompt photons detected by the 
experiment. A large-$p_T$ photon cross-section contains two contributions. In one of them the 
prompt photon is coupled to a quark in the hard subprocess. At leading order, 
since the partonic hard process is of the type $a + b \to c +\gamma$, the photon is not accompanied by another parton. 
These contributions are called the direct contributions. The second contribution
corresponds to a reaction with, in a first step, a parton emitted at large-$p_T$, followed 
by the Bremsstrahlung of a photon (essentially a collinear radiation). This second contribution is called the fragmentation contribution.

When a cone or a hollow-cone isolation is imposed on the cross section it cuts the two contributions in two different ways. It has
no effect on the LO Direct contribution, since the photon is never accompanied by another 
parton (only the direct HO contribution may be modified). On the contrary the fragmentation contribution is 
always modified since the photon is always accompanied by a  collinear parton.Therefore the dependence on 
the cone radius $R$ (or $r$) of the two contributions is very different~; the LO direct contribution is 
insensitive to $R$, but a $\ln R$ (or $\ln r$) appears in the HO direct contribution and in the fragmentation contribution.

A preceeding paper \cite{3r} was devoted to the detailed study of the cone-criterion and of the $R$ dependency 
of the cross section, a dependency which may be important, since small values of $R$ are
experimentally necessary at LHC in order to minimize the effect of the pile-up. We demonstrated that in the 
narrow cone case ($R \ll 1$) it is necessary to resum the $\ln R$ present in the 
cross-section. Otherwise the physical constraint $\sigma^{\text{inclusive}} > \sigma^{\text{cone}}$ ($\sigma^{\text{inclusive}}$ is the non isolated cross section),that we call unitarity, is violated. We also showed that the resummation of 
the large $\ln R$ (part of the HO of the direct component) leads to cross sections 
which no longer violate unitarity anymore, even at very small values of $R$. In our study we did not 
resum the logarithm contained in the fragmentation contribution which is very small because of isolation.

In the same paper we also studied a more realistic isolation, a ring isolation or hollow-cone
isolation, formed by two cones of radius $r$ and $R$, defining a ring in which the transverse hadronic energy is limited  
$E_{T \, \text{ring} }< E_{T \, \text{cut}}$.

We found that, after the resummation of the $\ln \, r$ of the direct  contribution, we were not able to verify the unitary condition. 
The reason is that the fragmentation contribution is now large since there is no more isolation in the small cone. 
Therefore this component must be studied more carefully and the $\ln \, r$ contributions that it contains also resummed. This is the aim of this paper.\\

The outline of the paper is the following. Section 2 is a reminder of the cone isolation of size $R$ and of the resummation of the $\ln R$ in the direct contribution. Then, in section 3, the hollow-cone isolation and $\ln r$ resummation in the direct component is discussed from the point of view of the unitarity. In section 4, we show how the resummation of the fragmentation contribution helps solve pratically the unitarity problem of the preceding section. We conclude in section 5.

\mysection{Cone isolation in the LL approximation}
\hspace*{\parindent}
The inclusive isolated distribution $d\sigma^{cone}/dp_T^\gamma d\eta^\gamma$ with transverse-energy isolation, which we simply denote by $\sigma^{cone} (p^\gamma ; z_c, R)$, is written as in eqn. (4.14) of ref. \cite{3r}~:
\bea
\label{2.1e}
\sigma^{\text{cone}} (p^\gamma;z_c,R)&=&\sum_a \int_0^1 {dz \over z} \, \widehat{\sigma}^{a, \text{cone}} \left ( {p^\gamma \over z} ; {z_c \over z}, R ; \mu , M, M_F \right )D_a^\gamma (z; M_F) \ \theta (z - z_c) \nn \\
&+&\widehat{\sigma}^{\gamma, \text{cone}} (p^\gamma ; z_c, R ; \mu , M, M_F) \ , 
\eea

\noi where
\beq
\label{2.2e}
z_c = {p_T^\gamma \over E_{T \, \text{cut}}+ p_T^\gamma}\ = {1\over 1+\epsilon},
\eeq
$M$ and $M_F$ are the factorisation scales of the initial and final distribution functions.
At small values of $R$, the partonic cross section of the direct component contains logarithmic terms of the type $\alpha_s^{m+1}(\alpha_s \, \ln R)^k$, and the partonic cross section of the fragmentation component contains logarithmic terms of the type $\alpha_s^{m+2}( \alpha_s \, \ln R)^k$. The leading logarithmic (LL) terms are those with $m = 0$ (and $k = 1,2,3, ...$).

The resummation of the LL terms (the subscript notation $[\ ]_{LL}$ denotes the LL accuracy) produces the following result \cite {3r}
\begin{align}
  \hspace{2em}&\hspace{-2em}\left [\sigma^{\text{iso}} (p^\gamma ;z_c,R)\right ]_{LL} \notag \\
  &= {\alpha_s (\mu ) \over \pi} \sigma_\gamma^{\text{Born}} (p^\gamma ; M) +\left
 ( {\alpha_s (\mu )\over \pi}\right )^2 \sum_{a} \int_0^1 {dz \over z} \sigma_a^{\text{Born}}\left ( {p^\gamma \over
 z} ; M\right ) D_{a}^{(0)}( z ; {\cal M}, R \, p_T^\gamma ) \notag \\
  &\quad {} + \left ( {\alpha_s (\mu )\over \pi}\right )^2 \sum_{a,b} \int_0^1 {dz \over z} \, \sigma_a^{Born}\left ( {p^\gamma \over z} ; M\right ) \notag \\
  &\qquad \qquad \qquad \qquad {} \times \int_z^1 {dx \over x} \ E_{ab}^{(0)} \left (  {z \over x} ; {\cal M}, R \, p_T^\gamma\right )D_b^{\gamma (0)}(x; R \, p_T^\gamma ) \ \theta (x - z_c)\ .  
\label{2.3e}
\end{align}
where ${\cal M}$ is a scale of order $p_T^\gamma$ and $M_F$ is set to $R \, p_T^\gamma$.

\noi The expression of the parton evolution operator $E_{ab}^{(0)} (z; {\cal M}, R  \, p_T^\gamma )$ is given in ref. \cite{3r}. The explicit expression of the fragmentation function $D_a^{(0)} (z;{\cal M}, R  \, p_T^\gamma )$ is
\beq
\label{2.4e}
D^{(0)}_a(z;{\cal M},R \, p_T^\gamma ) = \sum_b \int_{(R \, p_T^\gamma )^2}^{{\cal M}^2}{dk^2 \over k^2} \int_z^1 {dx \over x} \ E_{ab}^{(0)} \left (  {z \over x}  ;{\cal M} , k\right )  K_b^{(0)}(x) \ .
\eeq
where $K_b^{(0)}$ is the lowest order fragmentation function $K_b^{(0)} = {\alpha \over2\pi}
e_b^2 {(1 + (1-x)^2) \over x}$ where $e_b$ is the electric charge of the parton $b$ in unit of $e$.

The LL resummation of the $\ln R$ terms have been implemented in the program {\tt JetPhox}
\cite{4r,5r}. This implementation supplements the complete NLO result \cite{4r}, which has the exact dependence on $R$, with the summation of all the LL terms beyond the NLO. The complete NLO result is added to a ``subtracted version''of the LL formula (\ref{2.3e}). This subtracted version avoids double counting of perturbative terms. It is obtained by considering the LL formula and by explicitly subtracting from it the terms that are obtained by expanding the same formula up to the NLO. A detailed discussion of the resummation of the fragmentaton contribution (third term of expression (2.3)) is given in section 4.

We add a comment about the actual implementation of resummation  in {\tt JetPhox}. We note that, due to the isolation cut $x > z_c$, the contribution of the fragmentation function $D_b^{\gamma \, (0)} (x;R \, p_T^\gamma )$ is small, so that the impact of the resummed factor $E^{(0)}(z,{\cal M}, R \, p_T^\gamma)$ in the third term of the right-hand side of eq. (\ref{2.3e}) is not significant. For the sake of numerical simplicity, this resummed factor is replaced by its truncation at $O(\alpha_s)$~:
\beq
\label{2.5e}
E_{ab}^{(0)}(z/x,{\cal M},R \, p_T^\gamma )\to \delta(1-z/x)\delta_{ab}+ {\alpha_s \over 2\pi} \ P_{ab}^{(0)} ( z/x) \ln \left ( {{\cal M}^2 \over (R \, p_T^\gamma )^2}\right ) \ .
\eeq


The results obtained in \cite{3r} in the case of a cone-isolation are summarized in table 1. To
study small values of $R$ we turned to the LHC kinematics and we considered high values of
$p_T^\gamma$. We consider proton-proton collisions at $\sqrt{s} = 7$~TeV, $p_T^\gamma =
100$~GeV and $|\eta^\gamma | < 0.6$. The energy isolation parameter is $\varepsilon = 0.04$ (i.e.,
$z_{c} = 0.9615$ and $E_{T \, \text{cut}} = 4$~GeV) and the radius of the isolation cone is varied in the
range $0.06 \leq R \leq 0.5$. The parton distribution functions are the CTEQ6M \cite{6r} ones, and
the fragmentation functions the set II of the BFG distributions \cite{7r}. \par

The first two columns of table 1 show that the NLO cross section is very stable with respect to changes of the factorization scale $M_F$ when the photon is isolated. Here we note a violation of unitarity for small values of $R$ ($R \leq 0.1$). The rightmost column displays the effect of resummation. It is small, and it is of the order of 7~\% at $R = 0.06$. Unitarity is no more violated down to the very small ($R = 0.06$) value. This value of $R$ is however extreme~; the experimental isolation cones typically have a radius bigger than $0.3$ for which the effect of resummation is even smaller ($\leq 1\ \%$).

\begin{table}[htb]
\begin{center}
\begin{tabular}{|c|c|c|c|}
\hline
$R$ &NLO &NLO &NLO \\
&$p_T^\gamma /2$ &$R p_T^\gamma$ &$R p_T^\gamma$-resummed \\
 \hline
.5 &3.59 &3.59 &3.57  \\
.3 &3.86 &3.85 &3.81\\
.1 &4.35 &4.34 &4.19 \\
.06 &4.56 &4.55 &4.24 \\
\hline
{\small No isol} &4.29 & & \\
\hline
\end{tabular}
\caption{Variation with $R$ of the total cross sections (in pb/GeV) at $\sqrt{s} = 7$~TeV. The renormalisation scale $\mu$ and the factorization scale $M$ of the parton distribution functions are set equal to $p_T^\gamma/2$. For the photon fragmentation function we study the case $M_F = p_T^\gamma/2$ and $M_F = R \, p_T ^\gamma$}
\label{tab1}
\end{center}
\end{table}

\mysection{Implications for a hollow-cone criterion}
\hspace*{\parindent}
Using the standard cone criterion or an alternative one, such as the Frixione criterion \cite{8r}, \cite{8s})
or one of its discretized version \cite{9r} or else the so-called hybrid isolation criterion \cite{10r}, it may be experimentally difficult to apply a cut on the accompanying energy in the region where the electromagnetic shower develops in the detector, since it may be difficult to disentangle the accompanying energy from the photon energy inside that region.  The shape and size of this region are detector dependent. To first approximation, the electromagnetic shower roughly fills a cone of radius $r\sim 0.1$; considering the standard cone criterion, this would correspond to an inner narrow cone of inefficient isolation inside the usual cone of radius $R$. For a ``hollow'' cone with an inner (empty) cone of such a small size ($r \sim 0.1$), the issue of the resummation of $\ln r$ contributions can matter. 
Let us also remind that the hollow-cone criterion selects photons that are {\it less} isolated than those selected by the standard cone critertion. Therefore, at fixed values of $E_{T \, \text{cut}}$ for both criteria, the corresponding cross sections have to fulfill the physical requirement
\beq
\sigma^{\text{inclusive}} (p^\gamma ) \geq \sigma^{\text{hollow}} (p^\gamma ; R, r) \geq \sigma^{\text{cone}}(p^\gamma	 ; R) \ ,
\label{3.1e}
\eeq

We used {\tt Jetphox} to compute the NLO cross section for the hollow-cone criterion, and we present quantitative results by considering the LHC kinematical configuration already discussed. We consider a hollow-cone criterion with the inner cone of radius $r = 0.1$ and the other cone of radius $R = 0.4$. The results for the NLO cross sections are given in Table 2, where they are also compared with the corresponding standard-cone cross section\footnote{The resummatiom effect in not included in the NLO-direct contribution. It is very small; cf. table 1 at $R = .5$.} (with $R =0.4)$.

\begin{table}[htb]
\begin{center}
\begin{tabular}{|c|cc|cc|c|}
\hline
 {\bf cone}&\multicolumn{2}{|c|}{\bf direct} &\multicolumn{2}{|c|}{\bf fragmentation} &{\bf total}  \\
\cline{2-6}
${\bf type}$ &{\bf Born} &{\bf NLO} &{\bf Born} &{\bf NLO} &{\bf NLO} \\
\hline
{\rm standard} &2.08 &3.56 &.077 &.165 &3.73 \\
 {\rm hollow} &2.08 &3.09 &.91 &.43 &3.52 \\
\hline
\end{tabular}
\caption{NLO cross sections (in pb/GeV) at $\sqrt{s} = 7$~TeV, with $\varepsilon = 0.04$ and $M_F = r \, p_T ^\gamma$. Comparison of the isolated cross sections for the standard cone ($R = 0.4$ and $M_F = R \, p_T^\gamma$) vs. the hollow-cone ($R = 0.4, r= 0.1$ and $M_F = r \, p_T^\gamma$) criterion.}
\label{tab2}
\end{center}
\end{table}

The results of tables 2 show that the bound $\sigma^{\text{hollow}}(p^\gamma, R, r) \geq \sigma^{\text{cone}} (p^\gamma , R)$ is violated. A detailed discussion of these calculations can be found in ref. \cite{3r}. The behavior of the NLO fragmentation contribution can be deduced from formula (\ref{2.5e}) which leads to a very negative contribution
\beq
\label{3.2e}
\sigma_{HO}^{\text{frag}} \sim - {\alpha_s \over 2 \pi}\ \ln {R^2 \over r^2} \cdot \ln {1 \over \varepsilon}
\eeq

\noi when $r$ and $\varepsilon$ are small. As a result the NLO fragmentation cross section is small compared to the Born cross section, and $\sigma^{\text{hollow}}$ is smaller than $\sigma^{\text{cone}}$. This demonstrates that approximation (\ref{2.5e}) is no more valid~; the $\ln r$ must be resummed. 

\mysection{Resummation of the fragmentation contribution}
\hspace*{\parindent}
In this section we discuss the resummation of the logarithms of the cone radius which appear in the HO corrections of the fragmentation cross section. The resummed expression is written $E_{ab}^{(0)}$ in formula (\ref{2.3e})~; until now we used the approximation (\ref{2.5e}) in sections 2 and 3 in which we notice that this approximation was no more valid. As expression (\ref{2.5e}) is already contained in the HO expression that we use in the calculation of the NLO fragmentation cross section, we must subtract it from the resummed expression and the last term of formula
 (\ref{2.3e}) is now written

\begin{align}
 \text{Last term of eq.~(\ref{2.3e})} &= \left ( {\alpha_s({\mu}) \over \pi} \right )^2 \sum_a \int_0^1 {dz \over z} \ \sigma_a^{Born} \left ( {p^\gamma \over z} , M \right ) \overline{D}_a (z,z_c, {\cal M} , R \, p_T^\gamma )
 \label{4.1e}
\end{align}
where
\begin{align}
 \overline{D}_a (z,z_c, {\cal M} , R \, p_T^\gamma ) &= \sum_b \, \int_z^1 {dx \over x} \left[ E_{ab}^{(0)}  \left( {z \over x}, {\cal M}, R \, p_T ^\gamma \right)  - {\alpha_s \over 2 \pi} \ P_{ab}^{(0)} \left({z\over x} \right) \ln  \left( {{\cal M} \over Rp_T^\gamma } \right)^2  \right] \notag \\
 &\qquad \qquad \qquad {} \times D_b^\gamma (x, R \, p_T^\gamma ) \theta (x - z_c) \ .
 \label{4.10e}
\end{align}

This expression corresponds to the cone-isolation criterion. The holow-cone case is considered below. We have dropped the index $(0)$ in the fragmentation function as we use a NLO function in our calculations. 
The $\overline{D}_a(z,R\,p_T^\gamma)$ fragmentation function is calculated by taking the inverse Mellin transform of the product, in the Mellin space,
of the resummed evolution kernel and of the fragmentation function $D_b^\gamma(n,R\,p_T^\gamma)$. All these calculations are performed with the choice ${\cal M} = p_T^\gamma$.  

The convolution (\ref{4.10e}) is the sum of two terms \footnote{we drop the indices a and b}.
\bea 
\theta (z - z_{c}) \int_z^1 {du \over u}  \left ( E^{(0)}(u) - {\alpha_s \over 2 \pi} \, P^{(0)} (u) \ln \left ( {{\cal M} \over R \, p_T} \right )^2\right ) \,
  D\left ( {z \over u}\right ) \nn \\
+ \ \theta (z_{c} - z) \int_z^{z/z_c} {du \over u}  \left ( E^{(0)}(u) - {\alpha_s \over 2 \pi} \, P^{(0)}(u) \ln \left ( {{\cal M} \over Rp_T} \right )^2\right ) \, D\left ( {z \over u}\right ) \ .
\label{4.2e}
\eea

\noi The first one defined for $z > z_{c}$ is positive~; it contains contributions of distribution of the types $\delta (1 - u)$ and $1/(1- u)_+$. The second one for $z < z_{c}$ may be negative. We calculate the contribution of these two terms to the fragmentation isolated cross section (the last term of (\ref{2.3e})) using the parameters fixed below expression (\ref{2.5e}) and $R =.1$. With the condition $z > z_{c}$ we obtain $.106$ and $-.033$ with the condition $z_{c} > z$, thus a resummed fragmentation cross section equal to $.073$. The resummed evolution kernel is a LL resummation. To its contribution we add the HO contribution (containing ${\alpha_s \over 2\pi} \ P^{(0)} \ln \left ( {{\cal M} \over Rp_T}\right )^2$ subtracted in (\ref{4.10e})) which is equal to $.063$ leading to a resummed NLO cross section equal to $.136$. This result can be compared to the NLO (without LL resummation) cross section equal to $.118$. The results obtained with other values of $R$ are given in Table 3.

\begin{table}[htb]
\begin{center}
\begin{tabular}{|c|c|c|}
\hline
$R$ &Fragmentation-NLO &Fragmentation-NLO-resummed \\
 \hline
.5 &.173  &.203 \\
.3 &.157  &.181 \\
.1 &.118  &.136 \\
\hline
\end{tabular}
\caption{Comparison between unresummed and resummed fragmentation cross sections. The factorization scale is $R \, p_T^\gamma$.}
\label{tab3}
\end{center}
\end{table}

\noi The total direct + fragmentation cross section is increased by the resummation of the logarithm $\ln R$, but only slightly since the isolated fragmentation cross section is small.

 The case of the hollow-cone criteria is more interesting since the fragmentation cross section may play a more important role. The resummed hollow-cone cross section is written like the last term of expression (\ref{2.3e}), but with $E_{ab}^{(0)} ({z \over x} ; {\cal M}, R \, p_T^\gamma )$ $D_b^{\gamma} (x, R \, p_T^\gamma )$ replaced by
\beq
\label{4.3e}
E_{ab}^{(0)} ({z \over x} ; R \, p_T^\gamma , r \, p_T^\gamma ) D_b^{\gamma} (x, r \, p_T^\gamma ) \ .
\eeq

\noi The isolation is now in a ring of radii $r$ and $R$, with $r \ll R$, with the definition $z_{c} = {1 \over 1 + \varepsilon}$. We also introduce a condition on the hadronic energy accompanying the photon $E_{T \, \text{cut}}^{(r)} = \overline{\varepsilon} \,  p_T^\gamma$ in the cone of radius $r$ with the definition $v_{c} = 1/(1 + \overline{\varepsilon})$. The structure of the $z$-constraints on the integration in $z$ in formula (\ref{2.3e}) is now different from (\ref{4.2e}) and is written
\bea
&&\theta (z - z_{c}) \int_z^1 {du \over u} \left ( E^{(0)}(u) - {\alpha_s \over 2 \pi} \ P^{(0)}(u) \ln \left ({R \over r}\right )^2 \right ) D\left ({z \over u}\right ) + \nn \\
&&\theta (z_{c} - z)\ \theta (z - v_{c}) \int_{1 - z\varepsilon}^1 {du \over u} \left ( E^{(0)}(u) - {\alpha_s \over 2 \pi} \ P^{(0)}(u) \ln \left ({R \over r}\right )^2 \right ) D\left ({z \over u}\right ) + \nn \\
&&\theta (v_{c} - z)\ \theta (z - w_{c}) \int_{1 - z\varepsilon}^{z/v_{c}} {du \over u} \left ( E^{(0)}(u) - {\alpha_s \over 2 \pi} \ P^{(0)}(u) \ln \left ({R \over r}\right )^2 \right ) D\left ({z \over u}\right ) 
\label{4.4e}
\eea

\noi with the definition $w_{c} = {1 \over 1 + \varepsilon + \overline{\varepsilon}}$. When $\overline{\varepsilon} \to \infty$ (no isolation in the small cone) we obtain $v_{c} = 0$ and the $z$-structure of expression (\ref{4.4e}) almost reduces to that of expression (\ref{4.2e}). Let us separately consider the three contributions of expression (\ref{4.4e}). We use the same kinematical parameters as for the cone criterion with $R= 1.$ and $r = .1$. We also consider a slight isolation in the small cone $\overline{\varepsilon} = .5$.

With the first term of (\ref{4.4e}) we obtain the resummed NLO fragmentation contribution $d\sigma /dp_T^\gamma = .106$. The second term gives ${d\sigma \over dp_T^\gamma} = .561$ and the last $d\sigma /dp_T ^\gamma = - .0066$. These numbers lead to a LL resummed fragmentation cross section $d\sigma/dp_T^\gamma = .661$, and to the NLO resummed fragmentation cross section $d\sigma/dp_T^\gamma = .411$ (the HO term being equal to $-.25$). The total (direct + fragmentation) resummed NLO cross section is now $d\sigma /dp_T ^\gamma = 2.95$ and it must be compared to the cone-isolation results $d\sigma /dp_T^\gamma = 3.10$. The unitary condition  $\sigma^{\text{hollow}} > \sigma^{\text{iso}}$ is still violated, but less than in the unresummed case where $d\sigma^{\text{hollow}}/dp_T ^\gamma = 2.85$.

A similar comparison can be done with the values $R= .5$ and $r = .1, \, .2$. These results for the resummed direct and fragmentation cross sections are summarized in table 4

 \begin{table}[htb]
\begin{center}
\begin{tabular}{|c|c c c|c|ccc|}
  \hline
  $R$  &\hbox{$\scriptstyle{\rm Direct-Cone}$} &\hbox{$\scriptstyle{\rm Fragm.-Cone}$} &\hbox{$\scriptstyle{\rm Total-Cone}$}& $r$ &\hbox{$\scriptstyle{\rm Direct-hollow}$} &\hbox{$\scriptstyle{\rm Fragm.-hollow}$} &\hbox{$\scriptstyle{\rm Total-hollow}$} \\
 \hline
  1. &2.91 &.190 &3.10 &.1 &2.54 &.41 &2.95  \\
    \hline
    \multirow{2}{*}{.5} &\multirow{2}{*}{3.40} &\multirow{2}{*}{.204} &\multirow{2}{*}{3.60} &.1 &3.04 &.45 &3.49 \\
    \cline{5-8}
 & & & &.2 &3.03 &.61 &3.64 \\
    \hline
\end{tabular}
\caption{}Comparison between the cone isolation and the hollow-cone isolation.
\label{tab4}
\end{center}
\end{table}

We observe that a radius $R=.5$ improves the agreement with the unitarity condition. This latter is checked when the inner cone has a radius $r=.2$.
The numbers given above correspond to the case $\overline{\varepsilon} = .5$, a choice which enhances the direct and fragmentation cross section. We obtain (for $R = .5$) very similar values for $\overline{\varepsilon} = \infty$ (3.48) and  $\overline{\varepsilon} = .2 $ (3.51).
To sum up the cone isolation criterion leads to NLO-resummed cross-sections close to those obtained with the more "experimental" hollow-cone criterion.\\

However let us note that condition (3.1) has to be checked by physical cross sections corresponding to cross sections calculated at all orders in perturbation theory. In this paper we consider NLO cross-sections for which condition (3.1) is no more absolute. Thus violation of unitarity exhibited in Table 4 points towards higher order calculations.

\section{Conclusion}

The studies reported in this paper examine in details the effect of various isolation cuts on the large-$p_T$ photon cross sections. First we showed that the cone-criterion leads, at small values of R, to a violation of the unitarity condition ; this points toward the necessity of the resummation of the large $\ln R$. After resummation unitarity is checked.

At small values of $r$ (the radius of the inner cone) the hollow-cone criterion also leads to
a violation of unitarity. In this case the fragmentation contribution is important and the large $\ln r$ terms must be resummed. This resummation improves the situation, although unitarity is still weakly violated at small values of $r$ ($r \simeq . 1$) by the NLO-resummed cross-sections,
The final conclusion is that the inner cone of radius $r$ only slightly change the standard cone results, at least at NLO with the parameters used in this paper. Therefore calculations performed with the cone-criterion are sufficient to discribe the complex experimental isolation.

\section{Acknowledgements}
This work on the resummation of the large logarithms originating from the small size of the isolation cone has been started in collaboration with E. Pilon and S. Catani who both passed away after a long disease. 
We will always remember the pleasure of working with them, as well as their expertise and kindness.



\begin{thebibliography}{99} 
\bibitem{1r} Atlas coll., Phys.Rev. {\bf D83}(2011)052005
\bibitem{2r} CMS coll., Phys.Rev. {\bf D84}(2011)0520011 ; JHEP {\bf 01} (2012) 133
\bibitem{3r} S. Catani, M. Fontannaz, J. Ph. Guillet and E. Pilon, JHEP {\bf 1309} (2013) 007
\bibitem{4r} Z. Belghobsi, M. Fontannaz, J.Ph. Guillet, G. Heinrich, E. Pilon, M. Werlen, Phys.Rev. {\bf D79} (2009) 114024
\bibitem{5r} S. Catani, M. Fontannaz, J. Ph. Guillet and E. Pilon, JHEP {\bf 05} (2002) 028.
\bibitem{6r} J. Pumplin, D. R. Stump, J. Huston, H. L. Lai, P. M. Nadolsky and W. K. Tung, JHEP {\bf 0207} (2002) 012.
\bibitem{7r} L. Bourhis, M. Fontannaz and J. Ph. Guillet, Eur. Phys. J. {\bf C2} (1998) 529.
\bibitem{8r} S. Frixione, Phys. Lett. {\bf B429} (1998) 369.
\bibitem{8s}
E.~Hall and J.~Thaler,
JHEP \textbf{09} (2018), 164.
\bibitem{9r} G.~Belanger, F.~Boudjema, J.~P.~Guillet and E.~Pilon,\\
``Proceedings, 7th Les Houches Workshop on Physics at TeV Colliders: Les Houches, France, May 30-June 17, 2011''.
\bibitem{10r}
X.~Chen, T.~Gehrmann, N.~Glover, M.~H\"ofer and A.~Huss,
JHEP \textbf{04} (2020), 166.

\end{thebibliography}
  \end{document}